\input harvmac
\def\div{\nabla\!\cdot\!}
\def\curl{\nabla\times}

\def\tiny{\scriptscriptstyle\rm}

\def\half{{1\over 2}}

\def\kbT{k_{\scriptscriptstyle\rm B}T}

\def\dnb{\delta\vec n}
\def\ofx{({\bf x})}
\def\ofxp{({\bf x}')}
\def\ofxt{}

\def\bold#1{\setbox0=\hbox{$#1$}%
     \kern-.010em\copy0\kern-\wd0
     \kern.025em\copy0\kern-\wd0
     \kern-.020em\raise.0200em\box0 }

\def\brac#1#2{\{{#1},{#2}\}}
\def\thth#1{\Theta\left[z-s_{#1}\right]\Theta\left[e_{#1} -z\right]}

\lref\LARSON{L.A.~Archer and R.G.~Larson, J.~Chem.~Phys. {\bf 103}, 3108
(1995).}
\lref\PC{J.~Prost and N.A.~Clark, in {\sl Proceedings of the International
Conference on Liquid Crystals, Bangalore, 1979}, edited by S.~Chandrasekhar
(Heyden, Philadelphia, 1980); in {\sl Liquid Crystals of One and
Two-Dimensional
Order, Garmisch-Partenkirchen, Fanvar, 1980} edited by W.~Helfrich and
F.~Hepke (Springer, Berlin, 1980).}
\lref\RF{L.~Radzihovsky and E.~Frey, Phys. Rev. B {\bf 48}, 10357 (1993).}
\lref\FORI{D.~Forster, {\sl Hydrodynamic Fluctuations, Broken Symmetry,
and Correlation Functions} (W.A.~Benjamin, Reading, MA, 1975).}
\lref\FORII{D.~Forster, Annals of Physics {\bf 84} (1974) 505.}
\lref\FORIII{D.~Forster, Phys. Rev. Lett. {\bf 32} (1974) 1161.}
\lref\LUB{T.C.~Lubensky, Phys. Rev. A {\bf 2} (1970) 2497.}
\lref\KLN{P.~Le Doussal and D.R.~Nelson, Europhys. Lett. {\bf 15}, 161 (1991);
R.D. Kamien, P. Le~Doussal, and D.R.~Nelson, Phys. Rev. A {\bf
45}, 8727 (1992).}
\lref\SB{J.V.~Selinger and R.F.~Bruinsma, Phys. Rev. A {\bf 43}, 2910
(1991); Phys. Rev. A {\bf 43}, 2922 (1991).}
\lref\TONER{J.~Toner, Phys. Rev. Lett. {\bf 68}, 1331 (1992); R.D.~Kamien and
J.~Toner,
Phys. Rev. Lett. {\bf 74}, 3181 (1995).}
\lref\TARMEY{V.G.~Taratura and R.B.~Meyer, Liquid Crystals {\bf 2} 373
(1987).}
\lref\KN{R.D.~Kamien and D.R.~Nelson,
J. Stat. Phys. {\bf 71}, 23 (1993).}
\lref\DGP{P.G.~de Gennes and J.~Prost, {\sl The Physics of Liquid Crystals},
Second Edition, Chap. VII (Oxford University Press, New York, 1993).}
\lref\MORSE{D.C.~Morse, Phys. Rev. E {\bf 58}, 1237 (1998);
Macromolecules {\bf 31}, 7030 (1998); {\bf 31}, 7044 (1998).}
\lref\CL{P.M.~Chaikin and T.C.~Lubensky, {\sl Principles of Condensed Matter
Physics},
(Cambridge University Press, Cambridge, 1995).}
\lref\LWD{R.D.~Kamien, Eur. Phys. J. B {\bf 1}, 1 (1998).}
\lref\DE{M.~Doi and S.F.~Edwards,  {\sl The Theory of Polymer Dynamics},
(Oxford University Press, Oxford, 1986).}
\lref\RUB{A.N.~Semenov and M.~Rubinstein, Eur. Phys. B {\bf 1}, 87 (1998).}

\Title{}{\vbox{\centerline{Poisson Bracket Formulation of}\vskip 2pt\centerline
{Nematic Polymer Dynamics}}}

\centerline{Randall D. Kamien\footnote{$^\dagger$}{email: {\tt
kamien@physics.upenn.edu}}}
\smallskip\centerline{\sl Department of Physics and Astronomy, University of
Pennsylvania,
Philadelphia,
PA 19104}

\vskip .3in
We formulate the dynamical theory of nematic polymers, starting from a
microscopic
Poisson bracket approach.  We find that the Poisson bracket between the nematic
director and momentum depends on the (Maier-Saupe) order parameter
of the nematic phase.  We use this to derive reactive couplings
of the nematic director to the strain rates.  Additionally,
we find that local dynamics breaks down as the polymers begin to overlap.  We
offer
a physical picture for both results. 

\Date{21 June 1999; revised 29 June 1999}

\newsec{Introduction and Summary}
The dynamics of line-like objects, from biomolecules to flux-lines in
superconductors,
has become increasingly important to understand.   The processing and
manufacture of
petroleum products, food products and petroleum-food products requires the
control
of polymer flow.  The remarkable properties of superconductors can be exploited
only
if flux-flow dynamics is controlled.  While
the equilibrium statistical mechanics of line-like objects such as directed
polymers,
flux lines and dipole chains in ferro- and electrorheological fluids has been
formulated from the microscopic physics\refs{\SB,\KLN}, a similar theory
has not been developed for the dynamics.  While polymer entanglement
dramatically affects the
dynamics by introducing cosmologically long time scales, the statics is not
affected.
Indeed, it is the dynamics of entanglement that aids in pinning of
high-temperature
superconductors, in the strength of glue and in other remarkable rheological
properties.
Previously, dynamics have been formulated on a hydrodynamic basis, which
ignores the connectivity of the lines \RF .  Other approaches include effective
theories which use single-polymer response functions and neglect
polymer-polymer
interactions \MORSE .
Here, we study the dynamics of polymers which, due to
steric or other interactions, are already aligned into a nematic state.  By
considering a system which is orientationally ordered in equilibrium,
the effect of flow on alignment can be separated from the effect of
alignment on flow.

In this work we formulate the fluctuating
hydrodynamics of polymer nematics based on a microscopic,
Poisson bracket approach \FORI.  In this way we will account for polymer
degrees
of freedom as well as long-lived hydrodynamic variables.  Our principle result
is a fundamentally 
new derivation of the elusive reactive coupling $\lambda$ \FORIII\ between
the director field ${\bf n}$ and the velocity field $\bf v$:
\eqn\eelusive{
{\partial n_\mu\over \partial t} = \left[\delta_{\mu\nu} - n_\mu
n_\nu\right]\left\{{1+\lambda\over 2} n_\gamma\partial_\gamma v_\nu -
{1-\lambda\over 2}n_\gamma\partial_\nu v_\gamma\right\}.}
Unlike most parameters in hydrodynamics $\lambda$ is neither set by
Kubo formulae nor the fluctuation-dissipation theorem.  We will show
that to the order we work, our result agrees with the analysis
of Archer and Larson \LARSON\ which is based upon a phenomenological,
Landau-theory
dynamics \DE .  Thus our result both verifies the validity of both a microscopic
and a phenomenological approach.

As is typical of the former approach, we develop our theory in steps.  First we
identify the hydrodynamic variables.  In a polymer nematic system we have four
fields of interest: the areal polymer density $\rho\ofx$, the momentum density
${\bf g}\ofx$, the fluctuation of the nematic director $\dnb\ofx = {\bf n}\ofx
- {\bf n}_0$,
and the density of polymer ``heads'' and ``tails'', $\rho_{\tiny HT}\ofx$.
We then write these fields in terms of the positions and
momenta of the individual monomers.   We consider, as an example, the
areal density field $\rho\ofx$.  Since the polymers are directed we may write
the location of the $\alpha^{\rm th}$ polymer in terms of
an affine parameter $\tau_\alpha$ which marks the monomers:
\eqn\epolymplace{{\bf R}_\alpha(\tau_\alpha) = \left[r_\alpha^x(z)
+\tau_\alpha h_\alpha^x,
r_\alpha^y(z) + \tau_\alpha h_\alpha^y,\tau_\alpha+s_\alpha\right]}
where $\vec r_\alpha(z)$ is the (two-dimensional) displacement of the monomer
at height $z=\tau_\alpha+s_\alpha$
away from the ground state, straight-line configuration with the
two-dimensional
tangent $\vec h_\alpha$, and the polymers start
at
a height $z=s_\alpha$.  The expression for the areal density is:
\eqn\ebegi{\rho\ofx = \sum_\alpha \delta^2\left[\vec r_\alpha(z)
+ (z-s_\alpha)\vec h_\alpha- {\bf
x}_\perp\right]\thth{\alpha},}
where $e_\alpha$ is the height of the polymer end.
This expression requires further explanation and, at the same time, shows the
inherent
complexity of this problem.  The areal density is a sum of delta-functions in
the
plane at the positions of the polymers.  Moreover, since these polymers have
heads and tails
the product of Heaviside step functions $\Theta(\cdot)$ counts the polymer only
if we
consider heights between the start and end of the polymer at $z=s_\alpha$ and
$z=e_\alpha$
respectively.  We will investigate similar expressions for the remaining fields
in the
next section.

We note here that although expressions such as \epolymplace\ are awkward,
similar
expressions could not easily be formed in {\sl isotropic} polymer systems.  The
essential feature of polymer nematics is that chain contour length corresponds
to the common $z$-axis along which all the polymers are more or less aligned.
This
allows us to formulate our theory in space, so as to incorporate hydrodynamics,
while
at the same time including the microscopic polymer degrees of freedom.  These
degrees of freedom are accounted for via the Poisson brackets of the $\vec
r_\alpha(z)$
with the conjugate momentum $\vec p_\alpha(z)$.  From these, we can generate
the Poisson brackets of the hydrodynamic fields: if we
can get a closed set of brackets then we can construct a consistent
hydrodynamic
theory.  However, we see that the
areal density \ebegi\ not only depends on $\vec r_\alpha(z)$ but also on the
location of the polymer heads and tails as well as the average polymer
tangent $\vec h_\alpha$.  The head and tail locations are accounted for via
the field $\rho_{\tiny HT}\ofx$ described above.  What do we do with the
average polymer tangent $\vec h_\alpha$?  Though the nematic phase has
two Nambu-Goldstone modes associated with spontaneously-broken rotational
invariance, there is, in addition, a massive mode, the Maier-Saupe order
parameter $S$ which measures the degree of nematic ordering.  The $\vec
h_\alpha$
are associated with this mode.  The order parameter can change locally via
changes in the local polymer alignment.  Presumably, if the polymers
are entangled, the time scale for
these rearrangements corresponds to some sort of reptation time required for
the release of topological constraints.   On the other hand, if the polymers
are short (unentangled) then there is a separation of time scales
which allows us to approximately decouple the local Maier-Saupe mode.
In both these limits, we can decouple these
tilt modes from the monomer modes.  This will allow us to {\sl pre-average} the
tilt modes to arrive at effective Poisson brackets for the remaining fields.

The next step in our formulation of dynamics requires a coarse-grained
free energy in terms of our macroscopic variables.  In section III we will
propose a free energy consistent with the symmetries of the system which
also includes the equilibrium physics of the polymer ends.  From this
we will derive dynamical equations for the hydrodynamic variables.

\newsec{Definition of Hydrodynamic Variables}

We start by formulating the theory in terms of microscopic, directed
trajectories of the
individual polymers, $\{\vec r_\alpha(z)\}$, where $\alpha$ labels the polymer
and $\vec r(z)$ is the displacement of the polymer from its equilibrium
position.  Recall that the nematic state is characterized by the
Maier-Saupe order parameter,
\eqn\ems{S={3\langle\,\cos^2\theta\,\rangle -1\over 2}}
where $\langle\cdot\rangle$ denotes a thermal, ensemble average, and
$\theta$ is the angle between the nematic director and the
ordering axis (taken throughout to be $\hat z$).
When $S\neq 1$, the directors do not all line up along $\hat z$.  Since
the nematic order parameter is set by the details of the isotropic--nematic
transition, it is a massive, non-hydrodynamic mode. 
In
order
to take into account the ground state value of $S$ we
will assume that the equilibrium polymer
trajectory is ${\bf R_\alpha}(z) =
[(z-s_\alpha) h^x_\alpha,
(z-s_\alpha) h^y_\alpha, z]$ and that $\langle\,\vec
r_\alpha(z)\,\rangle=0$.
Thus, the microscopic director is $d\vec r/dz +\vec h$.  Finally,
the $\mu$ component of the momentum of
polymer $\alpha$ at $z$ is $p^\mu_\alpha(z)$.

Unlike previous formulations \LUB, we
do not consider the traceless symmetric tensor $Q_{ij}$ typically used
for nematics.  Rather, deep in the nematic state, we are only concentrating
on the two Goldstone modes of the broken rotational symmetry.  As we shall see
this leads to a {\sl closed} set of Poisson brackets to lowest
order in derivatives and the director fluctuation $\delta\vec n$.  
We will incorporate
the ``up-down'' symmetry
of nematics via the symmetry of the free energy, just as is done in the
usual equilibrium description of nematics in terms of the unit-vector director
field ${\bf\hat n}$.

\subsec{Canonical Variables and Separation of Time Scales}

While our description of the chain locations is adequate in terms
of $\vec r_\alpha(z)$ and $\vec h_\alpha$, we should take care
in finding correct canonical variables from which to construct
macroscopic Poisson brackets.
We imagine starting with a Hamiltonian which is a function the actual polymer
location ${\bf R}_\alpha(\tau_\alpha)$, as in \epolymplace , and the conjugate
momentum ${\bf p}_\alpha(\tau_\alpha)$.
The equations of motion in the $xy$-plane are
\eqn\eeom{\dot p^i_\alpha(\tau_\alpha) = -{\delta H\over \delta
R^i_\alpha(\tau_\alpha)}\qquad\qquad
\dot R^i(\tau_\alpha) = {\delta H\over \delta p^i_\alpha(\tau_\alpha)}}
If $\vec h_\alpha$ is constant then these
equations of motion are precisely
\eqn\eeomii{\eqalign{\dot p^i_\alpha(z)
&= \brac{p^i_\alpha(z)}{H}\cr
\dot r^i_\alpha(z) &= \brac{r^i_\alpha(z)}{H}\cr}}
with
\eqn\empbb{\brac{r^i_\alpha(z)}{r^j_\beta(z')} =
\brac{p^i_\alpha(z)}{p^j_\beta(z')} =0,}
and
\eqn\empbiib{\brac{r^i_\alpha(z)}{p^j_\beta(z')} =
\delta_{\alpha\beta}\delta^{ij}\delta
(z-z'),}
where $i$ and $j$ run over the $xy$-plane.  We have made the
identification ${\bf p}_\alpha(z) = {\bf p}_\alpha(\tau_\alpha)$ with $
z=\tau_\alpha+s_\alpha$.
We must also calculate
the Poisson bracket of $p^z_\alpha(z)$ with the spatial co\"ordinate.
Motion along the $z$-axis should
be accounted for in motion of the polymer ends.  Indeed, motion of
the monomers up or down depends only on the motion of the starting point
$s_\alpha$.  Thus
\eqn\empb{\brac{r^i_\alpha(z)}{p^z_\beta(z')}=0}
but
\eqn\empbii{\brac{z_\alpha}{p^z_\beta(z')} = \delta_{\alpha\beta}
\delta(z_\alpha-z'),}
where we use the symbol $z_\alpha$ to remind the reader that the commutator
does not vanish only for momenta and positions on the {\sl same} polymer.
In other words, $z_\alpha$ takes the place of the $z$-component of $\vec
r_i(z)$
and $z_\alpha=s_\alpha+\tau_\alpha$.  We remind the reader that the fact that the
affine parameter $\tau$ can be mapped one-to-one with the co\"ordinate
height $z$ is precisely the simplification which we are exploiting.  Of
course, when we consider hydrodynamic variables, $z_\alpha$ will just become
the height co\"ordinate $z$.

We must now justify our neglect of the dynamics of $\vec h_\alpha$.  As
we discussed in the introduction, we can do this in two limits.  One limit
is when the chains are long and highly-entangled.  In this case we expect
that rearrangements of the nematic texture require topological constraint
release
or, in other words, reptation.  We are concentrating here on dynamics
on time scales shorter than the reptation time.  We could extend this
work by including some sort of reptation-based dynamics for the tilt fields
$\vec h_\alpha$, thus incorporating both hydrodynamics and topological
constraints in the same theory.

There is another limit of interest.  If the polymers are short, we expect
that the modes associated with the Maier-Saupe order parameter
will be the slowest to relax to an equilibrium distribution.  In this case, we
expect
that
on the intermediate time scales we are focusing on, the distribution of $\vec
h_\alpha$ will also be static, but consistent with the background nematic
order.
In this case the tilt modes can be averaged over, thus
explicitly removing the dependence of the fields on them.  We
contrast this average to a {\sl thermodynamic} average:
in this case, the average is over an ensemble
of different frozen-in textures while in thermodynamics we take the time
average of a single molecule.  In both these cases we get the same mean-square
average of the tilt field, though the correlations will differ.

Thus, in both the very long polymer and short polymer case we can pre-average
the
tilt degrees of freedom.

\subsec{Macroscopic Fields}
We must first define the coarse-grained, macroscopic fields, namely the
mass density, the director and the momentum density.
The mass density is given by:
\eqna\emd
$${\rho\ofx
= \sum_\alpha \delta^2\left(\vec r_\alpha(z) + (z-s_\alpha)\vec h_\alpha- {\bf
x}_\perp\right)\thth{\alpha}}\eqno\hbox{$\emd a$}$$
As described in the introduction, the terms in the sum
come from the following considerations: 1) The Dirac delta
function ``counts'' the areal mass (note that it is a {\sl two-dimensional}
delta function, defined in the $xy$ plane). 2) The pair of Heaviside step
functions
($\Theta[\cdot]$) control the polymer lengths.  The $\alpha^{\rm th}$ polymer
begins at
$z=s_\alpha$ and ends at $z=e_\alpha$.

The remaining macroscopic fields are defined via
$$\eqalignno{
[\rho\delta n^i]\ofx &= \sum_\alpha
\left({\displaystyle{dr_\alpha^i(z)}\over dz} +
h_\alpha^i\right)
\delta^2\left(\vec r_\alpha(z)+ (z-s_\alpha)\vec h_\alpha - {\vec
x}_\perp\right){\thth{\alpha}
\over\sqrt{1+\vec h_\alpha^2}}\qquad&\emd b\cr
g^\mu\ofx&=\sum_\alpha p_\alpha^\mu(z)\delta^2\left(\vec r_\alpha(z)+ \vec
h_\alpha(z-s_\alpha)
- {\vec x}_\perp\right)\thth{\alpha}&
\emd c\cr}$$
Note that an additional factor of $\left(1+\vec h_\alpha^2\right)^{-1/2}$
arises
in the definition
of $\dnb\ofx$ since the normalized unit
(average) tangent vector is
\eqn\etangnorm{{\bf T}_\alpha={\left[h^x_\alpha,h^y_\alpha,1\right]\over
\sqrt{1+\vec h^2}}.}
Note that the na\"\i ve definitions of the macroscopic fields, {\sl e.g.}
\eqn\naive{\rho\ofx =\sum_\alpha \delta^2[\vec r_\alpha -{\vec
x}_\perp]\thth{\alpha},}
applies only
when $\vec h_\alpha\equiv \vec 0$, or, in other words, when the nematic ground
state has
Maier-Saupe order parameter, $S=1$.

\newsec{Pre-Averaging the Nematic Texture}
As we discussed in the introduction, the Poisson brackets of the
hydrodynamic fields cannot close since they all depend on the average
polymer tilt $\vec h_\alpha$.  As discussed in the preceding, we
would like to pre-average over these degrees of freedom to arrive at
Poisson brackets for the hydrodynamic fields.
This procedure can be justified in two limiting cases.  

Consider the dynamics of a single polymer of length $L$.  This can be
decomposed
into normal modes {\sl ala} Rouse or Zimm \DE .  In either case the
$n^{\rm th}$ mode has wavenumber $q_n=2\pi n/L$ with a relaxation frequency $\omega_n
\propto
n^2/L^2$.  For very short polymers the separation in time scales between the
lowest
mode $n=1$ and the other modes is large.  In this limit we can decouple the
longest mode -- the tilt mode -- characterized by $\vec h$.  We could thus
average
the tilts over the collection of polymers.  The mean-square of the tilt fields
is simply related to the Maier-Saupe order parameter $S$.  We make the
distinction between the time average of a single polymer's tilt field which
gives
equilibrium statistics and the quenched average over the frozen polymer tilts.

As we imagine lengthening the polymers, the splitting in timescales becomes
less
pronounced and their spectrum becomes continuous.  However, when the polymer is
sufficiently
long, a {\sl new} dynamics dominates the longest wavenumber modes: reptation
dynamics.
In this regime, the tilt modes can only relax via activated reptation \RUB\
which
is an especially apt description of the likely rearrangements in polymer
nematics.
Thus, when the polymer is long enough to be entangled we also have a large
separation of time scales.  As a result, in this limit, we may also treat the
polymer
tilt field as essentially frozen.

We have now two distinct averages: the first is an average over a massive
degree of freedom that sets the value of $S$.  In the present formulation, this
average is over $\vec h_\alpha$.  The second average is the thermal average
over
fluctuations about the ground state, {\sl i.e.} averages over $\vec
r_\alpha(z)$.
We will treat these two averages separately, which amounts to treating the
average over $\vec h_\alpha$ as a {\sl quenched} average.
To connect with the Maier-Saupe order parameter, we have
\eqn\emsh{S = {3\over 2} \overline{\left(1\over 1 +\vec h_\alpha^2\right)}
-{1\over 2}}
since $\cos\theta$ in \ems\ is just the $z$ component of the average
tangent vector and where we have denoted the quenched average of $X$ by
$\overline{X}$.
Expanding for small $\vert\vec h_\alpha\vert$, we have
${2\over 3}(1-S )\approx \overline{\vec h_\alpha^2}$.  We will then take
averages over $\vec h_\alpha$
weighted by
\eqn\epofh{P(\vec h_\alpha)= {1\over 2\pi\Delta}
e^{-\vert\vec h\vert^2/2\Delta}}
where $\Delta= {1\over 3}(1-S)$ so that \emsh\ is satisfied.

In order to find the average macroscopic field variables, we must regularize
the Dirac
delta functions appearing in \emd{} .
To do this we
represent them as Gaussians with
the widths taken to be the excluded area of each polymer in a fixed $z$-plane.
We define
\eqn\edist{\delta^2(\vec r;a) = {1\over{2\pi a}}e^{\displaystyle{-\vert
\vec r\,\vert^2/2a}}.}
Then, when we average over $\vec h_\alpha$ weighting by \epofh\ we find
\eqn\esmear{\overline{\delta^2[\vec r_\alpha(z) + (z-s_\alpha)\vec h_\alpha
- {\bf x}_\perp; a]} = \delta^2[\vec r_\alpha(z) -{\bf x}_\perp; a
+\Delta(z-s_\alpha)^2].
}

We can thus calculate the tilt-averaged fields.
Since we have only included terms for the tangent
vector in \emd{b}\ to leading order in $\vec h_\alpha$, we
calculate average field values to order $\Delta$:
\eqna\emdefhbar{
$$\eqalignno{
\bar\rho\ofx&=\sum_\alpha\delta^2\left(\vec r_\alpha(z)-{\bf
x}_\perp; a + \Delta(z-s_\alpha)^2\right)\Xi_\alpha(z)&\emdefhbar a\cr
[\bar\rho\bar\delta n^i]\ofx &=\sum_\alpha
\left(1-{\Delta}\right){\displaystyle{dr_\alpha^i(z)}\over dz}
\delta^2\left(\vec r_\alpha(z) -{\vec x}_\perp; a
+ \Delta(z-s_\alpha)^2\right)\Xi_\alpha(z) & ~~\cr
{}~&\qquad
-\sum_\alpha (z-s_\alpha){\Delta}\partial_i
\delta^2\left(\vec r_\alpha(z) -{\vec x}_\perp; a
+ \Delta(z-s_\alpha)^2\right)\Xi_\alpha(z)&\emdefhbar b\cr
\bar g^\mu\ofx&=\sum_\alpha p^\mu_\alpha(z)
\delta^2\left(\vec r_\alpha(z) -{\vec x}_\perp; a
+ \Delta(z-s_\alpha)^2\right)\Xi_\alpha(z)&\emdefhbar c\cr}$$}
where $\Xi_\alpha(z)\equiv\thth{\alpha}$.

\subsec{The Breakdown of Local Dynamics}

Typically, Poisson brackets for dynamical variables are of the
form
\eqn\etheform{\brac{\rho(x)}{\vec g(x')} =
-\vec\nabla_x\left[\rho(x)\delta^3(x-x')
\right].}
The presence of the delta-function in \etheform\ along with a
local free energy density leads to local dynamics \CL , {\sl i.e.} every term
in the dynamical equations is evaluated at the same point in space and time.
Such a result must, of course, be taken with a grain of salt.  The
delta function
is an idealization to a system composed of point-like constituents.  In
reality,
these delta functions should be replaced with a smeared-out distribution.
Usually, this presents no problem -- calculations may be done with the
smeared-out distribution and the limit may be taken at the end.  However,
this is not the case in this situation.

{}From \esmear , it is easy to see that we may
only replace the disorder smeared delta-functions with the original
distributions
(of width $a$) when $a\gg\Delta(z-s_\alpha)^2$.  In other words,
it is only when the original width is wider than the disorder-averaged width
that
the distributions are delta-function like.
Since the presence of the term
$\thth{\alpha}$ in the expressions for the hydrodynamic variables limits
the maximum value of $z$, we have
$\Delta(z-s_\alpha)^2<\Delta(e_\alpha-s_\alpha)^2 =
\Delta\ell^2$, where $\ell$ is the typical polymer length.  Thus
we see that the original distributions are recovered everywhere whenever
$\ell\sqrt{\overline{h^2}}\equiv\ell\sqrt{\Delta}\ll a$.  This is easy to
interpret; when the average wandering of the polymer away from its starting
point, $\ell\sqrt{\Delta}$
is smaller than the interpolymer spacing, we may continue to treat the polymers
and their interactions as local.  Once the polymers have tipped out of their
cages the dynamics will become, necessarily, non-local.  A force at ${\bf x}$
leads
to a reaction at ${\bf x}'$ if a {\sl single} polymer can go from $\bf x$ to
${\bf x}'$.  Presumably, the typical polymer length $\ell$ should be replaced
by $\ell_P$, the polymer persistence length, when $\ell>\ell_P$ and 
the locality condition becomes $\Delta \ell_P^2 \ll a^2$.

In this delocalized regime the formalism will break down.  In particular, the
Poisson brackets of the coarse-grained fields, calculated in terms of \empb\
and
\empbii\ will be straightforward but complicated, and will necessarily lead to
non-local dynamics, {\sl even if the free-energy density arises from local
interactions}.
In the Conclusion we will discuss possible alternatives
to a non-local formalism based on different sets of variables.

\newsec{Fluctuating Hydrodynamics}
We can, however, consider the case of $\Delta\ll 1$.  This could arise
in the description of flux lines in superconductors or polymer nematics
in applied fields where the ground state polymer configurations are
almost always parallel to $\hat z$.  In any of these limits, it
is appropriate to replace the spread out delta functions $\delta^2\left(\cdot;a
+
\Delta(z-s)^2\right)$ with point-like delta functions.  In this case, we can
calculate the Poisson brackets of the average fields in terms
of the canonical, microscopic brackets.  We find the following, non-zero
brackets,
to lowest order in derivatives (of both $\vec r_\alpha(z)$ and delta
functions) and $\delta n$:
\eqna\epb{$$\eqalignno{
\brac{\bar\rho\ofx}{\bar g^\mu\ofxp}&=
-\left[\delta^\mu_\nu -
\delta^\mu_z\delta^z_\nu\right]
\partial^\nu_x\left[\rho\ofx\delta^3({\vec x
-\vec x'})\right]&{}\cr &\qquad+ \left(1+{\Delta}\right)
\delta^\mu_z\left[\rho_{\tiny HT}\ofx\delta^3({\vec x -
\vec x'})-\partial_i\left(\rho\delta n^i\ofx\delta^3({\vec x-\vec
x'})\right)\right]&\epb a\cr
\brac{\bar\delta n^i\ofx}{\bar g^\mu\ofxp}&= -\left[\delta^\mu_\nu -
\delta^\mu_z\delta^z_\nu\right]\delta^3({\vec x-\vec x'})\partial^\nu_x\delta
n^i\ofx&{}\cr
&\qquad+ \left[\left(1-{\Delta}\right)\delta^{i\mu}\partial^z_x
- {\Delta}\delta^{\mu z}\partial^i_x\right]\delta^3({\vec x-\vec x'})
&\epb b\cr
\brac{\bar g^\mu\ofx}{\bar g^\nu\ofxp}&= -\partial_x^\nu\left[g^\mu\ofx
\delta^3({\vec x-\vec x'})\right] + \partial^\mu_{x'}\left[g^\nu\ofxp\delta^3({
\vec x-\vec x'})\right]&\epb c\cr
}$$}
We have been forced to introduce an additional hydrodynamic field, $\rho_{\tiny
HT}\ofx$, the
density of polymer heads and tails, defined
as
\eqn\erht{\rho_{\tiny HT}\ofx \equiv \sum_\alpha \left(1-{\Delta}\right)
\left\{\delta^2(\vec r_\alpha(s_\alpha)-{\vec x}_\perp)
\delta(z-s_\alpha)-\delta^2(\vec r_\alpha(e_\alpha)-{\vec x}_\perp)
\delta(e_\alpha-z)\right\}}
Note that the second term in \epb{a}\ is exactly equal to
$\partial_z\bar\rho\ofx$.
In other words, we have
\eqn\econser{\left(1-{\Delta}\right)\partial_z\bar\rho\ofx +
\partial_i\left[\bar\rho
\delta \bar n_i\ofx\right] = \rho_{\tiny HT}\ofx}
This constraint is the familiar constraint of line liquids \refs{\TARMEY,\KN}\
which conserves polymer number but for polymer ends.  The prefactor of $(1 -
{\Delta})$
comes from the average over the $z$ direction.  In other words, the
rotationally invariant
conservation law is $\vec n_0\cdot\nabla\rho + \nabla\cdot\vec n = \rho_{\tiny
HT}$.
When averaging over the intrinsic randomness of $\vec n_0$, we get a factor of
$(1-{\Delta})$ in front of $\partial_z\rho$.  It is not present in
other terms because they are higher order in derivatives and powers of $\delta\vec n$.
\subsec{Equations of Motion}

With these Poisson brackets we can systematically derive the reactive terms
of hydrodynamics.  The viscosities will appear in accord with the symmetries of
the system.  Being uniaxial there will be $5$ viscosities \DGP , characterized
by
the viscous stress tensor 
$V_{\alpha\beta}=\eta_{\alpha\beta\gamma\beta}A_{\gamma\beta}$ where
\eqn\evisc{\eqalign{\eta_{\mu\nu\gamma\beta} &=
\eta_1 \left[n_\mu n_\mu n_\gamma n_\beta \right]
+\eta_2\left[\delta^{\perp}_{\mu\gamma}
\delta^{\perp}_{\nu\beta} + \delta^{\perp}_{\mu\beta}\delta^{\perp}_{\nu\gamma}
-
\delta^{\perp}_{\mu\nu}\delta^{\perp}_{\gamma\beta}\right]
+ \eta_4\left[\delta^{\perp}_{\mu\nu}\delta^{\perp}_{\gamma\beta}\right]\cr
&+ \eta_3\left[n_\mu n_\gamma\delta^{\perp}_{\nu\beta} + n_\nu
n_\gamma\delta^{\perp}_{\mu\beta}
+ n_\mu n_\beta\delta^{\perp}_{\nu\gamma}+n_\nu
n_\gamma\delta^{\perp}_{\mu\beta}\right]
+ \eta_5\left[\delta^{\perp}_{\mu\nu}n_\gamma n_\beta + n_\mu n_\nu
\delta^{\perp}_{\gamma\beta}\right] \cr}}
where $\delta^{\perp}_{\mu\nu} = \delta_{\mu\nu} - n_\mu n_\nu$.
The form of $V$ is fixed by symmetries: by nematic inversion it must
be even in $\vec n$ and by the Kubo formulae \FORI\ it must be
symmetric under $\mu\leftrightarrow \nu$, $\gamma\leftrightarrow \beta$ and
$(\mu\nu)\leftrightarrow(\gamma\beta)$, and $A_{\gamma\beta} =
\half\left(\partial_\gamma
v_\beta + \partial_\beta v_\gamma\right)$ is the strain rate tensor constructed
out of the fluid velocity $\vec v$.   Expanding $\vec n = \hat z + \dnb$ we can
find
the viscosity tensor in the nematic phase.

The free energy from which we will derive the reactive terms is:
\eqn\efree{F'=\int d^3\!x\, \left\{ {\vec g^2\over 2\rho} + F_{\rm equil}
\right\}}
where the equilibrium free energy is the sum of terms:
\eqn\efreeii{F_{\rm equil}=F_{\dnb} + F_{\rm pol}}
and
\eqn\efreii{F_{\dnb} = \int d^3\!x\,\left\{  {K_1\over
2}\left(\div\dnb\right)^2 +
{K_2\over 2}\left(\curl\dnb\right)^2 +{K_3\over
2}\left(\partial_z\dnb\right)^2\right
\}}
is the usual Frank free energy for a nematic, while
\eqn\efreiii{
F_{\rm pol} = \int d^3\!x\,\left\{ {B\over 2}\delta\rho^2
+ {G\over 2}\left[\partial_z\delta\rho + \rho_0\div\dnb\right]^2\right\}}
is the free energy of the polymer \KLN .  In \efreiii\
$B$ is the two-dimensional bulk modulus, $\rho_0=\rho-\delta\rho$ is the
average, areal polymer density and $G$ is the fugacity for
hairpins and free ends.
We have replaced $\rho_{\tiny HT}$ with the expression in \econser , to
lowest order in the field fluctuations $\delta\rho$ and $\dnb$ (which is why
we drop the factor of $\left(1-\Delta\right)$ in front of
$\partial_z\delta\rho$).
The parameter $G$ is given by $G=\kbT\ell/2\rho_0$ where $\ell$ is the typical
polymer length \KLN .  Terms of the form $\delta\rho\div\dnb$ are not allowed
due to the ${\vec n}\rightarrow -\vec n$ symmetry of the nematic phase.

We are now able to derive the equations of motion for $\vec g$, $\dnb$ and
$\delta\rho$
(where we have dropped the bar over the variables denoting the $\vec h$
average).
They are (with $\vec v\ofxt=\vec g\ofxt/\rho\ofxt$):
\eqna\eeqom{$$\eqalignno{
\partial_t\rho\ofxt &= - \partial_i g_i\ofxt + \left(1+{\Delta}\right)
\rho_{\tiny HT}\ofxt v_z\ofxt - \partial_i\left(\delta n_i\ofxt g_z\ofxt\right)
&\eeqom a\cr
\partial_t\delta n_i\ofxt &= -\partial_j\left(v_j\ofxt\delta n_i\ofxt\right)
+\left(1-{\Delta}\right)\partial_z v_i\ofxt -{\Delta}\partial_i
v_z\ofxt
-\Gamma{\delta F[\delta n_i]\over \delta(\delta n_i\ofxt)} +
\theta_i\ofxt&\eeqom b\cr
\partial_t g_\mu\ofxt&= -\partial_\nu\left({g_\mu g_\nu\over\rho}\right) +
\left[\delta^\mu_\nu -
\delta^\mu_z\delta^z_\nu\right]\rho\partial_\nu\left({\delta F[\rho]\over
\delta\rho}\right) + \partial_\alpha V_{\alpha\mu} + \xi_\mu &\eeqom c\cr
}$$}
where, in order to approach thermodynamic equilibrium,
the noise terms $\xi_i(\vec x,t)$ and $\theta_i(\vec x,t)$ have the
following correlations \PC :
\eqna\eeqno{$$\eqalignno{
\langle\,\xi_\mu(\vec x,t)\xi_\beta(\vec x',t')\,\rangle & =
2\kbT\eta_{\mu\nu\gamma\beta}
\partial_\nu\partial_\gamma\delta^3(\vec x-\vec x')\delta(t-t')&\eeqno a\cr
\langle\,\theta_\mu(\vec x,t)\theta_\nu(\vec x' t')\,\rangle &=
2\kbT\Gamma\delta^{\perp}_{\mu\nu}
\delta^3(\vec x-\vec x')\delta(t-t')&\eeqno b\cr
\langle\,\xi_\mu(\vec x,t)\theta_\nu(\vec x',t')\,\rangle&=0&\eeqno c\cr}$$}

There are a number of features of these equations worth mention.  First, we
see that the typical polymer lengths come into the equations through the
coupling
$G$, present in the equations for $\partial_t\delta n_i$ and $\partial_t
g_\mu$.
When $G$ is large, corresponding to long polymers, we may take the
coefficient, $\rho_{\tiny HT}=0$.  In the short molecule limit, presumably
$G\approx 0$.  There is a crossover wavevector
$k_c=\sqrt{2B\rho_0^3/(k_{\scriptscriptstyle B}T\ell)}$ below which
the system behaves as a pure nematic and above which the length of the polymers
becomes important.  There is a corresponding timescale $\omega_c\propto
k_c^2\sim
1/\ell$ which delineates a similar short-to-long crossover.

In the short-polymer limit, care must be taken due to the nature of the
approximation
under which the Poisson brackets were derived.  In particular, since we treat
the polymer tangent field as a vector, we are, in principle, distinguishing
$\vec n$ and $-\vec n$.  However, for small fluctuations around equilibrium,
long polymer
nematics should not sample the entire space of fluctuations:  it is very
unlikely that
an entire polymer will rotate by $\pi$ around the $x$- or $y$-axis.  This
simplification
is what enabled us to get a closed set of Poisson brackets order by order
in a derivative expansion.  The effect of hairpins could be put in explicitly
through the introduction of a hairpin-density field \KLN.

\subsec{Reactive Couplings in the Nematodynamic Limit}
Finally, we note that \eeqom{b}\ gives an expression for the elusive reactive
parameter
$\lambda$ \FORI\ in nematodynamics.  By rotational invariance, $\vec
n\rightarrow -\vec n$
and the constraint
$\vec n^2=1$, the reactive part of the equation of motion
for $\vec n$ has the form:
\eqn\eforn{{dn_\mu\over dt} = \delta^{\perp}_{\mu\nu}
\left[{1+\lambda\over 2}n_\gamma\partial_\gamma v_\nu -
{1-\lambda\over 2}n_\gamma\partial_\nu
v_\gamma\right]}
We can thus identify
\eqn\ourlambda{\lambda = 1- 2\Delta= {1+2S\over 3}.}
This is in
general agreement with the results of Forster \FORIII\ who calculated
the value of $\lambda$ within a Poisson bracket formalism for $Q_{\mu\nu}$ the
symmetric, traceless order parameter for short molecule nematic liquid crystals
\DGP .  He found that
\eqn\efors{\lambda = {1 +2\alpha S\over 3},}
where $\alpha = (I_l + 2I_t)/(I_l - I_t)$ is a parameter depending on
the moments of inertia $I_l$ and $I_t$ of the nematogens parallel and
perpendicular to the nematic axis, respectively.  We thus recover
Forster's result when $\alpha=1$ or, in other words, when the aspect ratio
$I_l/I_t$
of the nematogens becomes infinite.  Presumably this is a consequence of
taking delta-function densities in the transverse plane.

It is interesting to compare this result to the work of Archer
and Larson \LARSON .  In the same limit of infinite aspect ratio,
they found an expression for $\lambda$ in terms
of the expectations of the second- and fourth-rank order parameters,
$\langle\,P_2\,\rangle\equiv S$ and $\langle\,P_4\,\rangle$ where $P_2$ and
$P_4$
are the second and fourth Legendre polynomials evaluated at
$x=\cos\theta$:
\eqn\elarson{\lambda = {15\langle\,P_2\,\rangle + 48\langle\,P_4\,\rangle +
42\over
105\langle\,P_2\,\rangle}.}
To be consistent, we should expand \elarson\ in powers of $\Delta=
(1-S)/3$ to compare with \ourlambda\ and compare linear terms.
First, to linear order
in $\Delta$,
\eqn\elinear{\langle\,\cos^4\theta\,\rangle
=1-2\langle\,\vec h_\alpha^2\,\rangle \approx \left(1-\langle\,\vec
h_\alpha^2\,\rangle\right)^2
\approx \langle\,\cos^2\theta\,\rangle^2} 
and so
$\langle\,P_4\,\rangle \approx (35S^2 -10S -7)/18$ which gives
\eqn\elarsonexpand{
\lambda = 1 + {2\over 3}(S-1) + {2\over 9}(S-1)^2 + \ldots .}
Thus to leading order in $(S-1)=3\Delta$ \elarson\ agrees with \ourlambda !
Thus within
the $S\approx 1$ limit our result should be consistent with the data as
in \LARSON .  We note, moreover, that since the isotropic-to-nematic
phase transition is first order, $S$ does not grow continuously from $0$.
Indeed,
in Maier-Saupe theory $S\approx 0.44$ at this transition \DGP , and thus
$\Delta\approx 0.2$.   Therefore the deviation between our result
and the more exact result \elarson\ should be small sufficiently well
aligned samples.  However, although \ourlambda\ may be quantitatively 
reasonable, it misses an essential qualitative feature: it is always less
than $1$ and thus predicts that nematics will always tumble.  

Though the linear result in $\Delta$ fails to
predict the crossover from flow aligning to tumbling behavior, 
the virtue of our derivation of $\lambda$ is that
we get a direct interpretation of its origin.  In a
highly aligned sample with $S=1$ only gradients of $v_\perp$ along the
$z$-direction
can lead to rotations of the molecule (see Figure 1).  On the other hand, when
$S<1$, gradients in the $x$-direction of $v_z$ can also rotate the molecules and
so the polymer nematics would always be in a tumbling mode.  A higher order
analysis in powers of $\Delta$ would be required to see if a $\lambda>1$ could
come from our direct approach.
 
\newsec{Conclusions}
In summary we have derived the Poisson brackets of the relevant degrees of
freedom for a polymer nematic.  We have shown that these canonical brackets
become highly non-local when the polymers start to overlap.  In the limit where
the polymers do not overlap, we have presented a microscopic 
derivation of the reactive coupling $\lambda$, a coupling which 
is not a long-wavelength limit of a correlation function.

Our analysis requires a ``locality condition'' in order to be
trustworthy.  While this may appear restrictive, one might imagine
that, at some level of coarse graining the polymers, the no overlap condition
might be met.  If we were to clump polymer regions correlated in the $xy$-plane
together and coarse grain to the scale of this ``entanglement correlation
length''
$\xi_e$, our analysis may be applicable.  Our analysis is {\sl especially}
applicable
to flux lines in superconductors.  There, $\rho_{\tiny HT}\equiv 0$ and there
is
no problem, even in principle, to considering the flux line tangents as {\sl
vectors}.
The flux lines, in addition, have $\Delta=0$, simplifying the theory further.
It is
perhaps in this context that one could hope to make the most theoretical
progress.

This derivation has shown the inherent, unavoidable non-locality in
polymer nematic dynamics.  An interesting possibility is to focus on
a different set of conserved or {\sl almost} conserved variables.  For
instance,
it may be possible to reformulate this dynamics in terms of an entanglement
density by calculating the Hopf density \LWD , a scalar which measures the
local curvature of the director configuration -- for long polymers curvature
is a measure of local entanglement.   This density could be used as
a starting point for a phenomenological theory of ``entanglement dynamics.''
Another variable of interest might be the repton density, which measures
length per unit length \ref\DGREP{P.-G. de Gennes,  J. Chem. Phys. {\bf 55},
572 (1971).}.
It may be possible to
develop a hydrodynamics for this field as well \ref\inp{R.D. Kamien,
{\sl unpublished} (1999).}.

\newsec{Acknowledgments}
It is a pleasure to acknowledge stimulating discussions with
D.~Forster, E.~Frey, T.~Lubensky,
S.T.~Milner, D.~Morse, R.~Pelcovits, T.~Powers, H.~Stark and J.~Toner.
This work was supported by NSF MRSEC Grant DMR96-32598, the Research
Corporation, the Donors of the Petroleum Research Fund,
administered by the American Chemical Society and the
Alfred P. Sloan Foundation.

\nfig\fone{Velocity fields and nematic directors with various gradients
and molecular orientations. a) and c) Gradient in the $z$-direction of
$v_\perp$.  This
will rotate the director for all values of $S$.  b) and d) Gradient in the
$x$-direction
of $v_z$.  Only in d) will this rotate the molecule, {\sl i.e.} when $S<1$.}

\listrefs
\listfigs

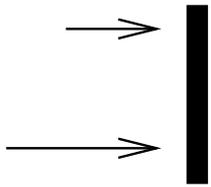
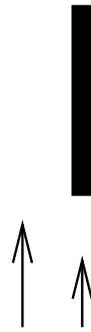
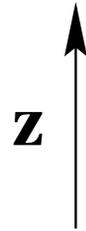
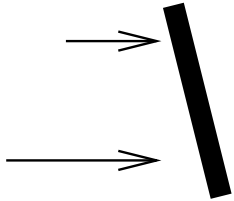
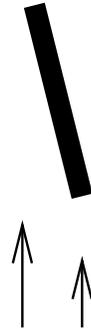
\bye